\documentstyle[aps,epsf]{revtex}

\begin{document}
\title{Exact dissipative cosmologies with stiff fluid}
\author{M. K. Mak\thanks{%
E-mail: mkmak@vtc.edu.hk}}
\address{Department of Physics, The Hong Kong University of Science and Technology,\\
Clear Water Bay, Hong Kong, P. R. China}
\author{T. Harko\thanks{%
E-mail: tcharko@hkusua.hku.hk}}
\address{Department of Physics, The University of Hong Kong,\\
Pokfulam, Hong Kong, P. R. China}
\maketitle

\begin{abstract}
The general solution of the gravitational field equations in the
flat Friedmann-Robertson-Walker geometry
is obtained in the framework of the full Israel-Stewart-Hiscock theory for a bulk
viscous stiff cosmological fluid, with bulk viscosity
coefficient proportional to the energy density.  

\end{abstract}

\section{Introduction}

Over thirty years ago Misner \cite{Mi66} suggested that the observed large
scale isotropy of the Universe is due to the action of the neutrino
viscosity which was effective when the Universe was about one second old.
Dissipative bulk viscous type thermodynamical processes are supposed to play
a crucial role in the dynamics and evolution of the early Universe. There
are many processes capable of producing bulk viscous stresses in the early
cosmological fluid like interaction between matter and radiation, quark and gluon
plasma viscosity, different components of dark matter, particle creation,
strings and topological defects etc. \cite{ChJa96}.

Traditionally, for the description of these phenomena the theories of Eckart \cite{Ec40} and Landau
and Lifshitz \cite{LaLi87} were used. But Israel \cite{Is76} and Israel and
Stewart \cite{IsSt76} have shown that the Eckart-type theories suffer from
serious drawbacks concerning causality and stability. Regardless of the
choice of equation of state, all equilibrium states in these theories are
unstable and in addition signals may be propagated through the fluid at
velocities exceeding the speed of light. These problems arise due to the
first order nature of the theory i.e. it considers only first-order
deviations from the equilibrium. The neglected second-order terms are
necessary to prevent non-causal and unstable behavior.

A relativistic consistent second-order theory was found by Israel \cite{Is76} and developed into what
is called transient or extended irreversible thermodynamics \cite{IsSt76}, 
\cite{HiLi89}, \cite{HiSa91}. Therefore, the best currently available theory
for analyzing dissipative processes in the Universe is the full Israel-Stewart-Hiscock causal
thermodynamics. Exact general solutions of the field equations for flat
homogeneous Universes filled with a full causal viscous fluid source for a
power-law dependence of the bulk viscosity coefficient on the the energy
density have been obtained recently in \cite{ChJa97}-\cite{MaHa98}.

It is the purpose of this Letter to obtain the general solution of the
gravitational field equations for the case of a bulk viscous cosmological
fluid, with the bulk viscosity coefficient proportional to the energy
density, obeying the Zeldovich (stiff) equation of state. In this case the
solution can be presented in an exact parametric form. The behavior of the
scale factor, deceleration parameter, viscous pressure, viscous pressure-thermodynamic pressure
ratio, comoving entropy and Ricci and Kretschmann invariants is also considered.

\section{Geometry, field equations and consequences}

For a Friedmann-Robertson-Walker (FRW) Universe with a line element $%
ds^{2}=dt^{2}-a^{2}(t)\left( dx^{2}+dy^{2}+dz^{2}\right) $ filled with a
bulk viscous cosmological fluid the energy-momentum tensor is given by 
\begin{equation}  \label{1}
T_{i}^{k}=\left( \rho +p+\Pi \right) u_{i}u^{k}-\left( p+\Pi \right) \delta
_{i}^{k},
\end{equation}
where $\rho $ is the energy density, $p$ the thermodynamic pressure, $\Pi $
the bulk viscous pressure and $u_{i}$ the four velocity satisfying the
condition $u_{i}u^{i}=1$. We use units so that $8\pi G=c=1$.

The gravitational field equations together with the continuity equation $%
T_{i;k}^{k}=0$ imply 
\begin{equation}  \label{field}
3H^{2}=\rho , 2\dot{H}+3H^{2}=-p-\Pi , \dot{\rho}+3\left( \rho +p\right)
H=-3H\Pi ,
\end{equation}
where $H=\dot{a}/a$ is the Hubble parameter. The causal evolution equation
for the bulk viscous pressure is given by \cite{Ma95} 
\begin{equation}  \label{bulk}
\tau \dot{\Pi}+\Pi =-3\xi H-\frac{1}{2}\tau \Pi \left( 3H+\frac{\dot{\tau}}{%
\tau }-\frac{\dot{\xi}}{\xi }-\frac{\dot{T}}{T}\right) ,
\end{equation}
where $T$ is the temperature, $\xi $ the bulk viscosity coefficient and $%
\tau $ the relaxation time. Eq. (\ref{bulk}) arises as the simplest way
(linear in $\Pi $) to satisfy the $H$ theorem ( i.e., for the entropy
production to be non-negative, $S_{;i}^{i}=\Pi ^{2}/\xi T\geq 0$, where $%
S^{i}=eN^{i}-\frac{\tau \Pi ^{2}}{2\xi T}u^{i}$ is the entropy flow vector, $%
e$ is the entropy per particle and $N^{i}=nu^{i}$ is the particle flow
vector) \cite{IsSt76}, \cite{HiLi89}.

In order to close the system of equations (\ref{field}) we have to give the
equation of state for $p$ and specify $T$, $\xi $ and $\tau $. As usual, we
assume the following phenomenological laws \cite{Ma95}: 
\begin{equation}  \label{csi}
p=\left( \gamma -1\right) \rho ,\xi =\alpha \rho ^{s},T=\beta \rho ^{r},\tau
=\xi \rho ^{-1}=\alpha \rho ^{s-1},
\end{equation}
where $1\leq \gamma \leq 2$ , and $\alpha \geq 0$, $\beta \geq 0$, $r\geq 0$
and $s\geq 0$ are constants. Eqs. (\ref{csi}) are standard in cosmological
models whereas the equation for $\tau $ is a simple procedure to ensure that
the speed of viscous pulses does not exceed the speed of light.

The requirement that the entropy is a state function imposes in the present
model the constraint $r=\left( \gamma -1\right) /\gamma $ \cite{ChJa97} so
that $0\leq r\leq 1/2$ for $1\leq \gamma \leq 2$.

The growth of the total comoving entropy $\Sigma $ over a proper time
interval $\left( t_{0},t\right) $ is given by \cite{Ma95}: 
\begin{equation}\label{ent}
\Sigma (t)-\Sigma \left( t_{0}\right) =-3k_{B}^{-1}\int_{t_{0}}^{t}\Pi
Ha^{3}T^{-1}dt,  
\end{equation}
where $k_{B}$ is the Boltzmanns constant.

The Israel-Stewart-Hiscock theory is derived under the assumption that the thermodynamical
state of the fluid is close to equilibrium, that is the non-equilibrium bulk
viscous pressure should be small when compared to the local equilibrium
pressure $\left| \Pi \right| <<p=\left( \gamma -1\right) \rho $ \cite{Zi96}.
If this condition is violated then one is effectively assuming that the
linear theory holds also in the nonlinear regime far from equilibrium. For a
fluid description of the matter , the condition ought to be satisfied.

To see if a cosmological model inflates or not it is convenient to introduce
the deceleration parameter $q=\frac{dH^{-1}}{dt}-1=\frac{\rho +3p+3\Pi }{%
2\rho }$. The positive sign of the deceleration parameter corresponds to
standard decelerating models whereas the negative sign indicates inflation.

With these assumptions the evolution equation for flat homogeneous causal
bulk viscous cosmological models is 
\begin{equation}  \label{ev}
\ddot{H}+3H\dot{H}+3^{1-s}\alpha ^{-1}H^{2-2s}\dot{H}-(1+r)H^{-1}\dot{H}^{2}+%
\frac{9}{4}\left( \gamma -2\right) H^{3}+\frac{3^{2-s}}{2\alpha }\gamma
H^{4-2s}=0.
\end{equation}

By introducing a set of non-dimensional variables $h$ and $\theta $ by means
of the transformations $H=\alpha _{0}h,t=\frac{2}{3\alpha _{0}}\theta $,
with $\alpha _{0}=\left( \frac{3^{s}\alpha }{2}\right) ^{\frac{1}{1-2s}%
},s\neq \frac{1}{2}$ and using the expression of $r$ as a function of $%
\gamma $, Eq. (\ref{ev}) takes the form 
\begin{equation}  \label{evol}
\frac{d^{2}h}{d\theta ^{2}}+\left[ 2h+h^{2\left( 1-s\right) }\right] \frac{dh%
}{d\theta }-\left( 1+r\right) h^{-1}\left( \frac{dh}{d\theta }\right) ^{2}+%
\frac{2r-1}{1-r}h^{3}+\frac{1}{1-r}h^{2\left( 2-s\right) }=0.
\end{equation}

By denoting $n=\left( 1-2s\right) /\left( 1-r\right) $ and changing the
variables according to $h=y^{1/\left( 1-r\right) },\eta =\int y^{1/^{\left(
1-r\right) }}d\theta $ Eq. (\ref{evol}) becomes: 
\begin{equation}  \label{final}
\frac{d^{2}y}{d\eta ^{2}}+\left( 2+y^{n}\right) \frac{dy}{d\eta }+\left(
2r-1\right) y+y^{n+1}=0.
\end{equation}

\section{General solution for a stiff cosmological fluid}

We consider the case of a stiff cosmological fluid with equilibrium pressure
equal to the energy density by taking the values of the parameters $\gamma
=2 $ and $r=1/2$. By introducing the substitutions $v=1/u$ and $u=dy/d\eta $
, Eq. (\ref{final}) can be transformed to a second type Abel first order
differential equation: 
\begin{equation}  \label{final1}
\frac{dv}{dy}=\left( 2+y^{n}\right) v^{2}+y^{n+1}v^{3}.
\end{equation}

A particular solution of Eq. (\ref{final1}) for $n=1$, $s=1/4$ has been
obtained in \cite{MaHa00}. We consider now another case of a stiff
cosmological fluid with bulk viscosity coefficient $\xi $ linearly
proportional to the energy density $\rho ,\xi =\alpha \rho $. Consequently $%
s=1$ and $n=-2$ and hence Eq. (\ref{final1}) becomes 
\begin{equation}\label{finx}
\frac{dv}{dy}=\frac{v^{3}}{y}+\left( 2+\frac{1}{y^{2}}\right) v^{2}=-\frac{%
v^{3}}{B(y)}-\left[ \frac{d}{dy}\frac{A(y)}{B(y)}\right] v^{2},
\end{equation}
where $A(y)=2y^{2}-1$ and $B(y)=-y$. By introducing a new variable $\sigma =%
\frac{1}{v}-\frac{A(y)}{B(y)}$, Eq. (\ref{finx}) can be written in the
general form 
\begin{equation}
\frac{dy}{d\sigma }=\sigma B(y)+A(y),
\end{equation}
or 
\begin{equation}\label{finy}
\frac{dy}{d\sigma }=2y^{2}-\sigma y-1.
\end{equation}

Hence we have transformed the initial Abel type equation into a Riccati
differential equation. A particular solution of Eq. (\ref{finy})is given by 
\begin{equation}
y=-\frac{\sigma }{1+\sigma ^{2}},
\end{equation}
and therefore the general solution of Eq. (\ref{finy}) is 
\begin{equation}
y\left( \sigma \right) =-\sigma \Delta \left( \sigma \right) +\frac{\Delta
^{2}\left( \sigma \right) e^{-\frac{\sigma ^{2}}{2}}}{C-2\int \Delta
^{2}\left( \sigma \right) e^{-\frac{\sigma ^{2}}{2}}d\sigma },
\end{equation}
where $\Delta \left( \sigma \right) =\left( 1+\sigma ^{2}\right) ^{-1}$ and $%
C$ is a constant of integration.

Hence the general solution of the gravitational field equations for a
Zeldovich causal bulk viscous fluid filled flat FRW Universe, with bulk
viscosity coefficient proportional to the energy density, can be obtained in
the following exact parametric form, with $\sigma $ taken as parameter: 
\begin{equation}\label{sol1}
t\left( \sigma \right) -t_{0}=-\frac{2}{3\alpha _{0}}\int \frac{d\sigma }{%
y(\sigma )},H\left( \sigma \right) =\alpha _{0}y^{2}\left( \sigma \right)
,a(\sigma )=a_{0}\exp \left[ -\frac{2}{3}\int y\left( \sigma \right) d\sigma %
\right], \rho \left( \sigma \right) =3\alpha _{0}^{2}y^{4}\left( \sigma \right),  
\end{equation}
\begin{equation}
q\left( \sigma \right) =5-\frac{3}{y^{2}\left( \sigma \right) }-\frac{%
3\sigma }{y\left( \sigma \right) },\Pi \left( \sigma \right) =6\alpha
_{0}^{2}y^{2}\left( \sigma \right) \left[ y^{2}\left( \sigma \right) -\sigma
y\left( \sigma \right) -1\right] ,\left| \frac{\Pi }{p}\right| =2\left| 1-%
\frac{1}{y^{2}\left( \sigma \right) }-\frac{\sigma }{y\left( \sigma \right) }%
\right|,
\end{equation}
\begin{equation}\label{sol2}
\Sigma \left( \sigma \right) =\Sigma \left( \sigma _{0}\right) +\frac{4\sqrt{%
3}\alpha _{0}a_{0}^{3}}{k_{B}\beta }\int y\left( \sigma \right) \left[
y^{2}\left( \sigma \right) -\sigma y\left( \sigma \right) -1\right] \exp %
\left[ -2\int y\left( \sigma \right) d\sigma \right] d\sigma,  
\end{equation}
where $a_{0}$, $t_{0}$ and $\Sigma \left( \sigma _{0}\right) $ are constant
of integration and $\alpha _{0}=2/3\alpha $.

The singular or non-singular character of the solution for all times $t\geq 0
$ can be checked from the finite (infinite) character of the Ricci invariant 
$R_{ij}R^{ij}$ and Kretschmann scalar $R_{ijkl}R^{ijkl}$, given by 
\begin{equation}
I=R_{ij}R^{ij}=12\alpha _{0}^{4}y^{4}\left[ \left( 3+3\sigma y-5y^{2}\right)
\left( 3+3\sigma y-4y^{2}\right) +y^{4}\right] ,
\end{equation}
\begin{equation}
J=R_{ijkl}R^{ijkl}=12\alpha _{0}^{4}y^{4}\left[ \left( 3+3\sigma
y-5y^{2}\right) ^{2}+y^{4}\right] .
\end{equation}

\section{Discussions and final remarks}

The evolution of the causal bulk viscous Zeldovich fluid filled  flat
Universe starts generally its evolution from a singular state, as can be
seen from the singular behavior of the invariants $I$ and $J$, presented,
for different values of the integration constant $C$, in Fig.1.
Generally, the dynamics of the Universe depends on the numerical values of $C$.

\begin{figure}
\epsfxsize=10cm
\centerline{\epsffile{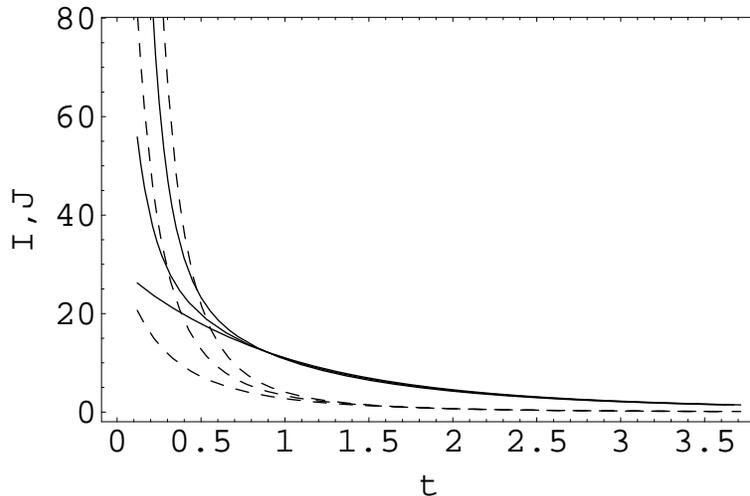}}
\caption{Time evolution of the invariants $I=R_{ij}R^{ij}$ (solid curves)
and $J=R_{ijkl}R^{ijkl}$ (dashed curves) for different values of the
integration constant $C$ ($\alpha _{0}=3/2$).}
\label{FIG1}
\end{figure}

The behavior of the scale factor is presented in Fig.2, for some specific values of the
integration constant. The evolution is expansionary for all times.
At the initial moment the scale factor is zero, while the energy density
tends to infinity.

\begin{figure}
\epsfxsize=10cm
\centerline{\epsffile{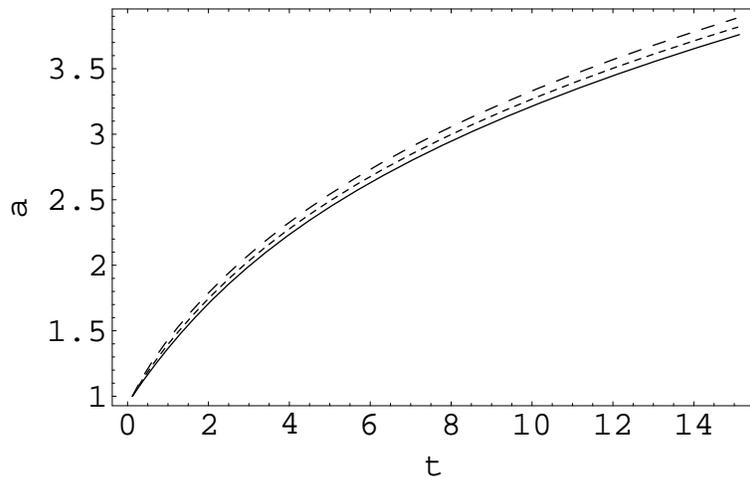}}
\caption{Time variation of the scale factor $a$ for different values of the
integration constant $C$: $C=-1.2$ (solid curve), $C=-1.1$ (dotted curve) and
$C=-1$ (dashed curve) ($\alpha _{0}=3/2$).}
\label{FIG2}
\end{figure}

The dynamics of the deceleration parameter, shown in Fig.3, indicates, for the chosen range
of the integration constant, a non-inflationary behavior for all times, with $q>0,\forall t\geq 0$.

\begin{figure}
\epsfxsize=10cm
\centerline{\epsffile{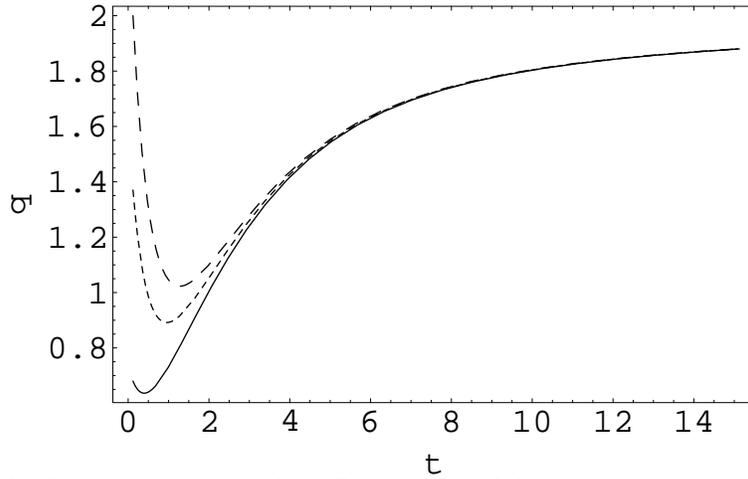}}
\caption{Evolution of the deceleration parameter $q$ for different values of the
integration constant $C$: $C=-1.2$ (solid curve), $C=-1.1$ (dotted curve) and
$C=-1$ (dashed curve) ($\alpha _{0}=3/2$).}
\label{FIG3}
\end{figure}

The bulk viscous pressure $\Pi $ is negative during the cosmological evolution, $%
\Pi <0,\forall t\geq 0$, as expected from a thermodynamic point of view. In
the large time limit, as can be seen from Fig. 4, the viscous pressure tends
to zero. In the same limit the bulk viscosity coefficient also becomes negligible small.

\begin{figure}
\epsfxsize=10cm
\centerline{\epsffile{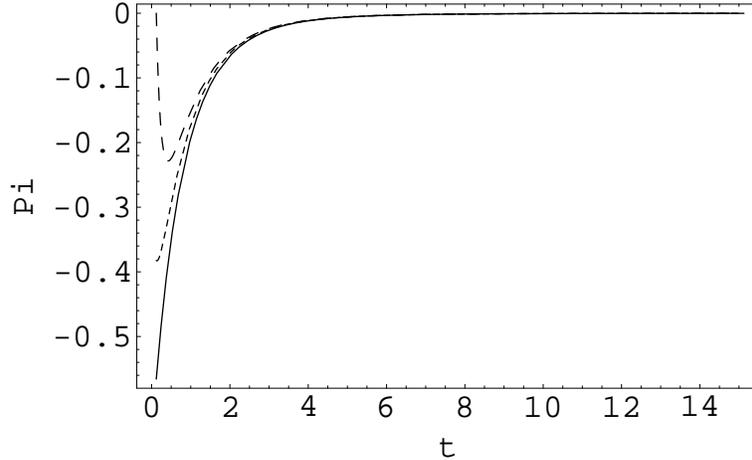}}
\caption{Time behavior of the bulk viscous pressure $\Pi $ for different values of the
integration constant $C$: $C=-1.2$ (solid curve), $C=-1.1$ (dotted curve) and
$C=-1$ (dashed curve) ($\alpha _{0}=3/2$).}
\label{FIG4}
\end{figure}

The ratio $l$ of the bulk and thermodynamic pressures, $l=\left|
\Pi /p\right| $ is presented in Fig. 5. For all times the condition $l<1$
holds and therefore the model is thermodynamically consistent.

\begin{figure}
\epsfxsize=10cm
\centerline{\epsffile{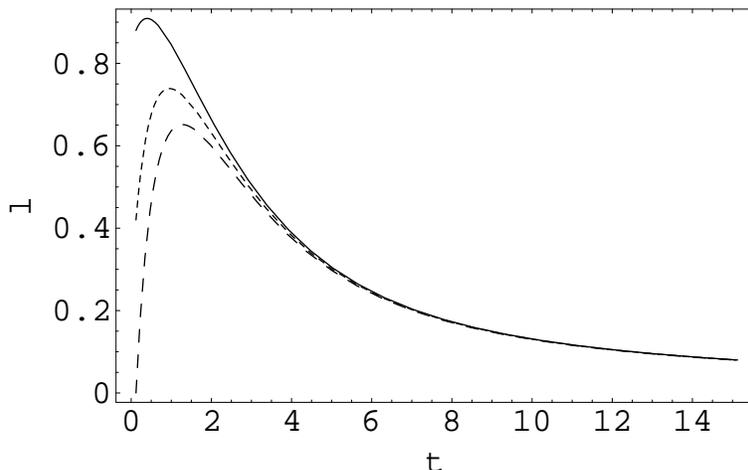}}
\caption{Variation with time of the ratio $l=\frac{\Pi }{p}$ for different values of the
integration constant $C$: $C=-1.2$ (solid curve), $C=-1.1$ (dotted curve) and
$C=-1$ (dashed curve) ($\alpha _{0}=3/2$).}
\label{FIG5}
\end{figure}

During the
cosmological evolution a large amount of comoving entropy is produced, with
the entropy $\Sigma $ increasing in time and tending in the large time limit
to a constant value.

\begin{figure}
\epsfxsize=10cm
\centerline{\epsffile{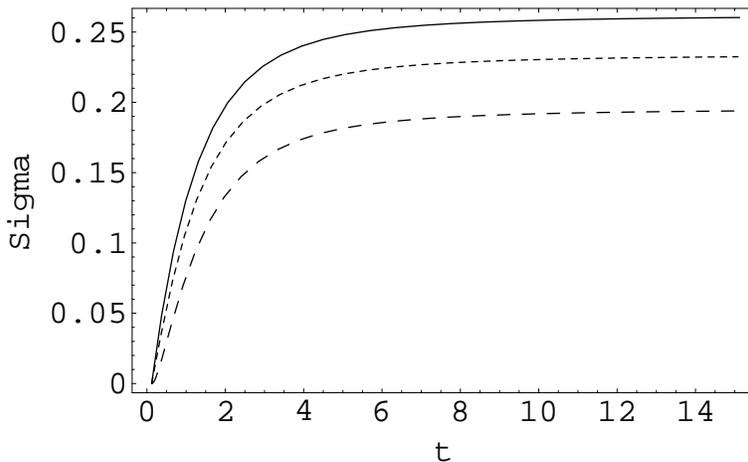}}
\caption{Dynamics of the comoving entropy $\Sigma $ for different values of the
integration constant $C$: $C=-1.2$ (solid curve), $C=-1.1$ (dotted curve) and
$C=-1$ (dashed curve). We have used the normalizations  $\alpha _{0}=3/2$ and
$\frac{4\sqrt{3}\alpha _{0}a_{0}^{3}}{k_{B}\beta }=1$.}
\label{FIG6}
\end{figure}

Due to the specific equation of state satisfied by the
bulk viscosity coefficient, the relaxation time $\tau $ is a constant in the
present model.

Most of the known exact solutions of the gravitational field equations with
a causal bulk viscous cosmological fluid do not satisfy the required
conditions of the thermodynamic consistency, leading to an inflationary
behavior and violating the condition of smallness of bulk viscous pressure. The solution
represented by Eqs. (\ref{sol1})-(\ref{sol2}) can
consistently describe the early dynamics of a superdense post-inflationary
era in the evolution of the matter dominated Universe, when, as expected, the bulk viscous
dissipative effects play an important role.

\end{document}